\documentclass[10pt,conference]{ieeeconf} 

\usepackage{setspace}
\usepackage{psfrag,subfigure}
\usepackage{amssymb, amsmath, cite}
\usepackage{mathrsfs}
\usepackage{color}
\usepackage{verbatim}
\usepackage{mathrsfs}
\usepackage{graphicx}
\usepackage{algorithm}
\usepackage{algpseudocode}
\usepackage{stmaryrd}

\usepackage[top=1in, bottom=0.7in, left=0.7in, right=0.7in]{geometry}
\usepackage{mathtools}
\usepackage{url}
\usepackage{pdfpages}
\usepackage{cite}

\usepackage{multicol}

\let\labelindent\relax
\usepackage{enumitem}
\setlist[description]{leftmargin=0.3cm,labelindent=0cm}

\newtheorem{assum}{Assumption}[section]
\newtheorem{remark}{Remark}[section]
\newtheorem{problem}{Problem}[section]

\IEEEoverridecommandlockouts 

\newcommand{\RR}{{\mathbb R}}

\newcommand{\cAOTS}{{\mathcal A}_{\textsc{\tiny OTS}}}
\newcommand{\cAPA}{{\mathcal A}_{\textsc{\tiny PA}}}

\newcommand{\cG}{{\mathcal G}}

\newcommand{\cO}{{\mathcal O}}

\newcommand{\cP}{{\mathcal P}}

\newcommand{\cX}{{\mathcal X}}

\newcommand{\cHMA}{{\mathcal H}_{\textsc{\tiny MA}}}
\newcommand{\EMA}{E_{\textsc{\tiny MA}}}
\newcommand{\XMA}{X_{\textsc{\tiny MA}}}
\newcommand{\IMA}{I_{\textsc{\tiny MA}}}
\newcommand{\GMA}{G_{\textsc{\tiny MA}}}
\newcommand{\RMA}{R_{\textsc{\tiny MA}}}
\newcommand{\LOTS}{L_{\textsc{\tiny OTS}}}
\newcommand{\EOTS}{E_{\textsc{\tiny OTS}}}
\newcommand{\EPA}{E_{\textsc{\tiny PA}}}
\newcommand{\DPA}{D_{\textsc{\tiny PA}}}
\newcommand{\HPA}{H_{\textsc{\tiny PA}}}

\newcommand{\QPA}{Q_{\textsc{\tiny PA}}}
\newcommand{\QMA}{Q_{\textsc{\tiny MA}}}

\newcommand{\sH}{\mathscr{H}}
\newcommand{\sF}{\mathscr{F}}
\newcommand{\sB}{\mathscr{B}}

\newcommand\tqed{\leavevmode\unskip\penalty9999 \hbox{}\nobreak\hfill\quad\hbox{$\triangleleft$}}

\title{\Large \bf 
A Framework for Multi-Vehicle Navigation using Feedback-Based Motion Primitives}
 
\author{Marijan Vukosavljev, Zachary Kroeze, Mireille E. Broucke, and Angela P. Schoellig
\thanks{Marijan Vukosavljev, Zachary Kroeze and Mireille E. Broucke are with the Dept. of Electrical and Computer Engineering, University of Toronto, Canada (e-mails: mario.vukosavljev@mail.utoronto.ca, zach.kroeze@mail.utoronto.ca, broucke@control.utoronto.ca). Angela P. Schoellig is with the University of Toronto Institute for Aerospace Studies (UTIAS), Canada (email: schoellig@utias.utoronto.ca). Supported by the Natural Sciences and Engineering Research Council of Canada (NSERC). }%
\thanks{Associated video at: http://tiny.cc/quad5scenes.}
}

\usepackage{fancyhdr}
\newcommand{\mytitle}{\textbf{Accepted final version.}
To appear in \textit{Proc. of the IEEE/RSJ International Conference on Intelligent Robots and Systems, 2017}.\\
\copyright 2017 IEEE. Personal use of this material is permitted. Permission from IEEE must be obtained for all other uses.}
\begin{document}


\maketitle

\thispagestyle{fancy}  
\pagestyle{empty}

\begin{abstract}

We present a hybrid control framework for solving a motion planning problem among a collection of heterogenous agents. The proposed approach utilizes a finite set of low-level {\em motion primitives}, each based on a {\em piecewise affine feedback control}, to generate complex motions in a gridded workspace. The constraints on allowable sequences of successive motion primitives are formalized through a {\em maneuver automaton}. At the higher level, a control policy generated by a {\em shortest path non-deterministic algorithm} determines which motion primitive is executed in each box of the gridded workspace. The overall framework yields a highly robust control design on both the low and high levels. We experimentally demonstrate the efficacy and robustness of this framework for multiple quadrocopters maneuvering in a 2D or 3D workspace.

\end{abstract}

\section{Introduction}
\label{sec:intro}

This paper presents a modular, hierarchical framework for motion planning and control of robotic systems.
Motion planning has received a great deal of attention by many researchers, and many hierarchical 
frameworks have already been presented.
We begin by reviewing the important features of existing frameworks and then we highlight the unique properties of our approach.

First, a common control specification is called {\em reach-avoid} \cite{LAV09,AYAN10,SATT14}, in which 
robotic agents must navigate an environment cluttered with obstacles to reach a final goal configuration.
More complex tasks expressed using {\em temporal logic} have also been considered \cite{BEL08,PAP09}.
Second, the physical workspace is typically {\em abstracted} into a finite number of regions
\cite{CHOS03,LAV09,KUM14} or points \cite{MIH09}. Planning at the high-level is often based on applying a 
{\em shortest path algorithm} to the graph arising from the workspace abstraction.
At the low-level, the control system is often assumed to be feedback linearizable or differentially flat 
\cite{MUR12,SCHW14} so that standard {\em trajectory tracking} \cite{FRAZ05,KUM14, SATT14} or 
{\em feedback control} methods \cite{HVS06,BEL08,AYAN10,GAZ14,CHOS03,PAP09,DIM13} can be 
applied. Recently the idea of {\em motion primitives} has further simplified the complexity of 
low-level control design \cite{DIM13,KUM14, SATT14}. Finally, feasibility of successive motion 
primitives can be formalized through a so-called {\em maneuver automaton} \cite{FRAZ05}. 

\begin{figure}[t]%
\centering%
\includegraphics[width=1\linewidth]{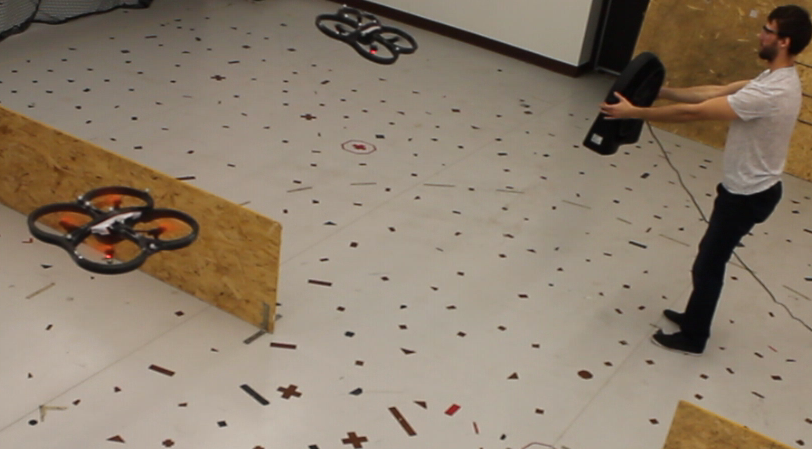}%
\caption{Two quadrocopters navigating around an obstacle in the presence of wind disturbances from a fan. A video illustrating the results can be found at http://tiny.cc/quad5scenes.}%
\vspace{-2mm}
\label{fig:expsetup}%
\end{figure}

In this paper we present a modular, hierarchical framework for motion control which incorporates 
the following features: 
(i) we consider a reach-avoid control specification; 
(ii) we abstract the workspace by partitioning it into rectangular regions; 
(iii) we assume the nonlinear control system has a translational symmetry in the output;
(iv) we perform high-level motion planning using standard shortest path algorithms;
(v) we employ a maneuver automaton to encode feasible sequences of motion primitives;
(vi) we perform low-level control design of motion primitives based on reach control theory \cite{RB06,HVS06}. 
We have selected these features because they provide an appealing balance between applicability and efficiency.

Despite the wealth of available literature, our framework offers important advantages over existing approaches, particularly due to the following three features: 

{\bf Robustness and safety.} \
Our framework uses low-level controllers based on {\em reach control theory}. It yields piecewise affine feedbacks that guarantee safety constraints on the states and outputs of the closed-loop system \cite{RB06, HVS06} without the need to specify feasible open-loop trajectories.
While other closed-loop methods such as potential methods \cite{CHOS03} can be used in our framework, reach control is used for its guarantees on safety and for its ease of design. 

{\bf Asynchronous motion of multiple agents.} \
In a multi-agent system, each agent may execute a motion primitive independently of the actions of other agents. This freedom of action is modelled in our framework in terms of the {\em non-determinism} of a graph which captures feasible sequences of motion primitives of the collection of agents. Although each robot's trajectory is deterministic for a given initial condition, the graph abstracts low-level state information regarding specific initial conditions so that the order of completion of motion primitives arises as non-determinism. This feature allows for time-independent asynchronous motion, which contrasts with many existing approaches where robot motion is typically synchronized or restricted to move one at a time \cite{BEL08,AYAN10, GAZ14}.

{\bf Computational feasibility.} \
One of our main modelling assumptions is that the system dynamics have a {\em translational symmetry} in the outputs \cite{FRAZ05}. This assumption implies that motion primitives can be first designed over a single box in the output space, and then be reapplied to other boxes since they are translations of each other. This feature significantly reduces the complexity of the high-level planning problem, allowing for a computationally feasible solution.


Finally, we highlight the differences between this paper and our previous work \cite{VUK16}. 
In \cite{VUK16}, reach control was demonstrated on a single quadrocopter exhibiting a one-dimensional side-to-side motion. The difficult steps of performing a state space triangulation of the safety region and specifying a high-level sequence on this triangulation were done manually. In this paper, we have addressed these difficulties by introducing features (i)-(vi). This enabled us to extend the reach control design in \cite{VUK16} to higher dimensional systems, which in this paper is demonstrated on multiple quadrocopters moving in several degrees of freedom simultaneously.

\section{Problem Statement} \label{sec:prob}

Consider the general nonlinear control system 
\begin{equation} 
\label{eq:nonlinsys}
\dot{x} = f(x,u), \;\;\;\;\; y = h(x),
\end{equation}
where $x \in \RR^n$ is the state, $u \in \RR^{\mu}$ is the control input, and $y \in \RR^p$ is the output, with $p \leq n$. This model can represent the aggregate dynamics of a multi-robot system with the states as the positions and velocities, and the outputs as the positions, possibly after feedback linearization or differential flatness transformations. We denote $x(t)$ as the state trajectory and $y(t)$ as the output trajectory of \eqref{eq:nonlinsys} under some initial condition and control.

This paper considers a reach-avoid problem to find a feedback control causing the output trajectories of \eqref{eq:nonlinsys} to reach a goal set while avoiding an obstacle set. Let $\cP \subset \RR^p$ denote a feasible output space. Let $\cO \subset \cP$ denote a subset of obstacle regions, and
let $\cG \subset \cP$ be a nonempty goal set. See Figure \ref{fig:OTSsquares} for an example. We consider the following problem.

\begin{problem}
\label{prob:reachavoid}
Given the system \eqref{eq:nonlinsys}, output space $\cP \subset \RR^p$, obstacle set $\cO \subset \cP$, and goal set $\cG \subset \cP$, find a feedback control strategy $u(x)$ and a set of initial conditions $\cX_0 \subset \RR^n$ such that
for each $x(0) \in \cX_0$ we have 
\begin{itemize}
\item[(i)]
{\bf Avoid}: 
for all $t \geq 0$, $y(t) \in \cP \setminus \cO$. 
\item[(ii)]
{\bf Reach}:  
there exists $T \geq 0$ such that for all $t \geq T$, $y(t) \in \cG$. 
\end{itemize}
\end{problem}

We make a standing assumption regarding the outputs of the system  \eqref{eq:nonlinsys} in order to exploit symmetry; see \cite{FRAZ05} for an exposition on nonlinear control systems with symmetries. This assumption is satisfied for many robotic systems, for example, when the outputs are positions. Also, it significantly simplifies our control design; see Section \ref{sec:MAexample}.

\begin{assum}
\label{assump:outputs}
For simplicity, the outputs are the first $p$ states, that is, $y_i = x_i$ for $i = 1,\ldots,p$. Moreover, the system has a {\em translational invariance} with respect to its outputs, that is, for all $u\in \RR^m$ and $x, x'\in \RR^n$ with $x_j'=x_j$, $j=p+1,\ldots,n$, we have $f(x,u) = f(x', u)$.
\tqed
\end{assum}

\section{Methodology} \label{sec:method}

\begin{figure}
\centering%
\includegraphics[width=1\linewidth,trim=0cm 0cm 0cm 0cm, clip=true]{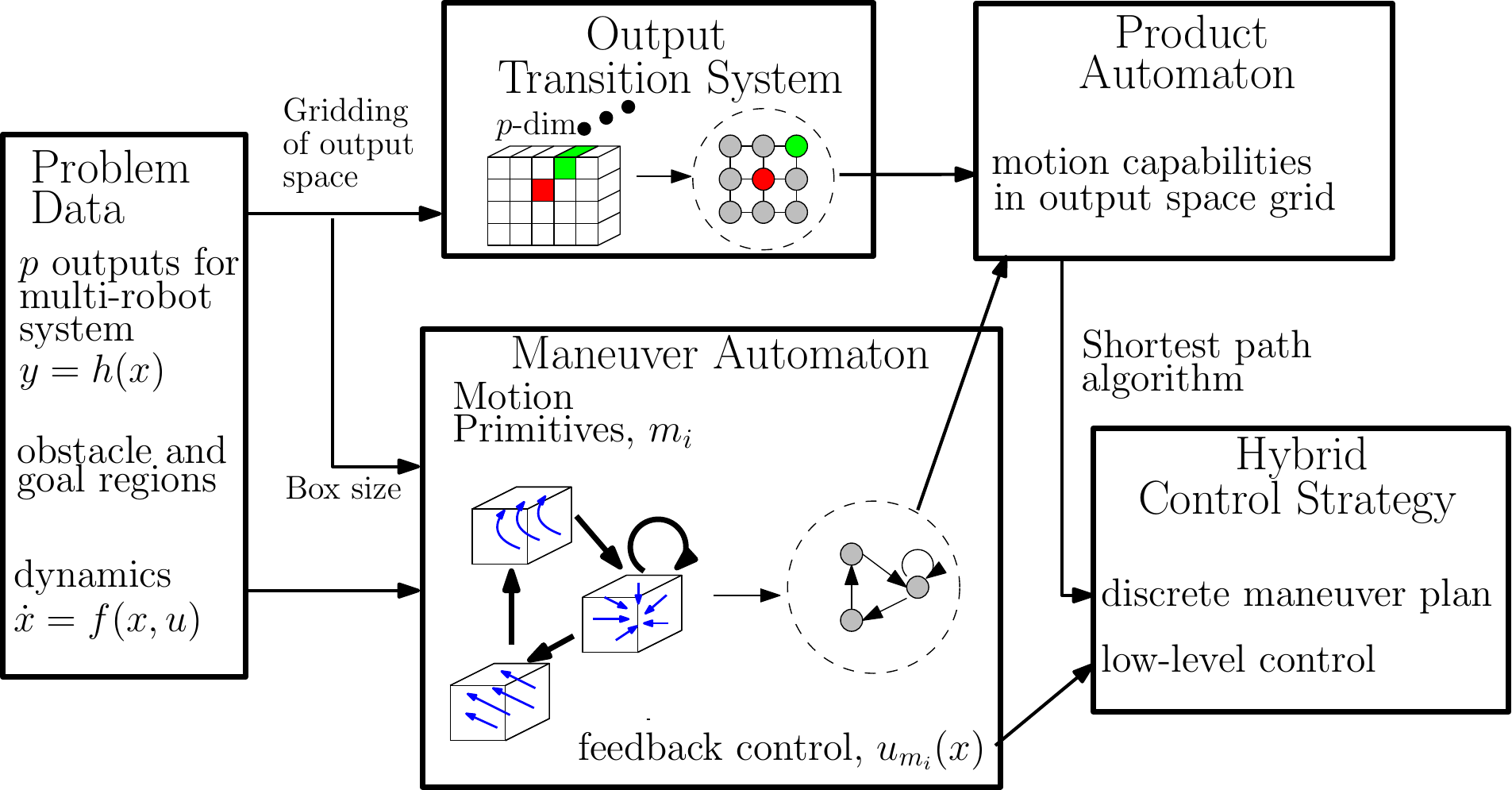}
\caption{Our hybrid control framework consists of five modules.
}%
\vspace{-2mm}
\label{fig:methodology}%
\end{figure}

In this section we present our methodology. It can be broken down into five main modules, as depicted in 
Figure~\ref{fig:methodology}. 
\begin{itemize}
\item The {\em problem data} include the system \eqref{eq:nonlinsys} with $p$ outputs satisfying Assumption \ref{assump:outputs}
and a reach-avoid task to be executed in the output space. 
\item The {\em output transition system} (OTS) is a graph that captures an abstraction of
the output space based on a partition of the output space into boxes. The reach-avoid task is translated in terms of desired behaviors on the OTS. 
\item The {\em maneuver automaton} (MA) is a hybrid system whose nodes correspond to all possible control modes, called {\em motion primitives}, that can be implemented over any single box in the output space. The edges of the MA represent feasible successive motion primitives.
\item The {\em product automaton} (PA) is a graph which is the synchronous product of the OTS and the discrete 
part of the MA. It represents the combined constraints on feasible motions in the output space and feasible successive motion primitives. 
\item The {\em hybrid control strategy} is a combination of the high-level plan and the low-level controllers obtained from the design of motion primitives. The high-level plan is obtained by applying a shortest path algorithm adapted to non-deterministic graphs on the PA \cite{BRO05}.
\end{itemize}

Now we describe in detail each of these modules. 

\subsection{Output Transition System} \label{sec:OTS}

\begin{figure}
\centering%
\includegraphics[width=1.0\linewidth,trim=0cm 0cm 0cm 0cm, clip=true]{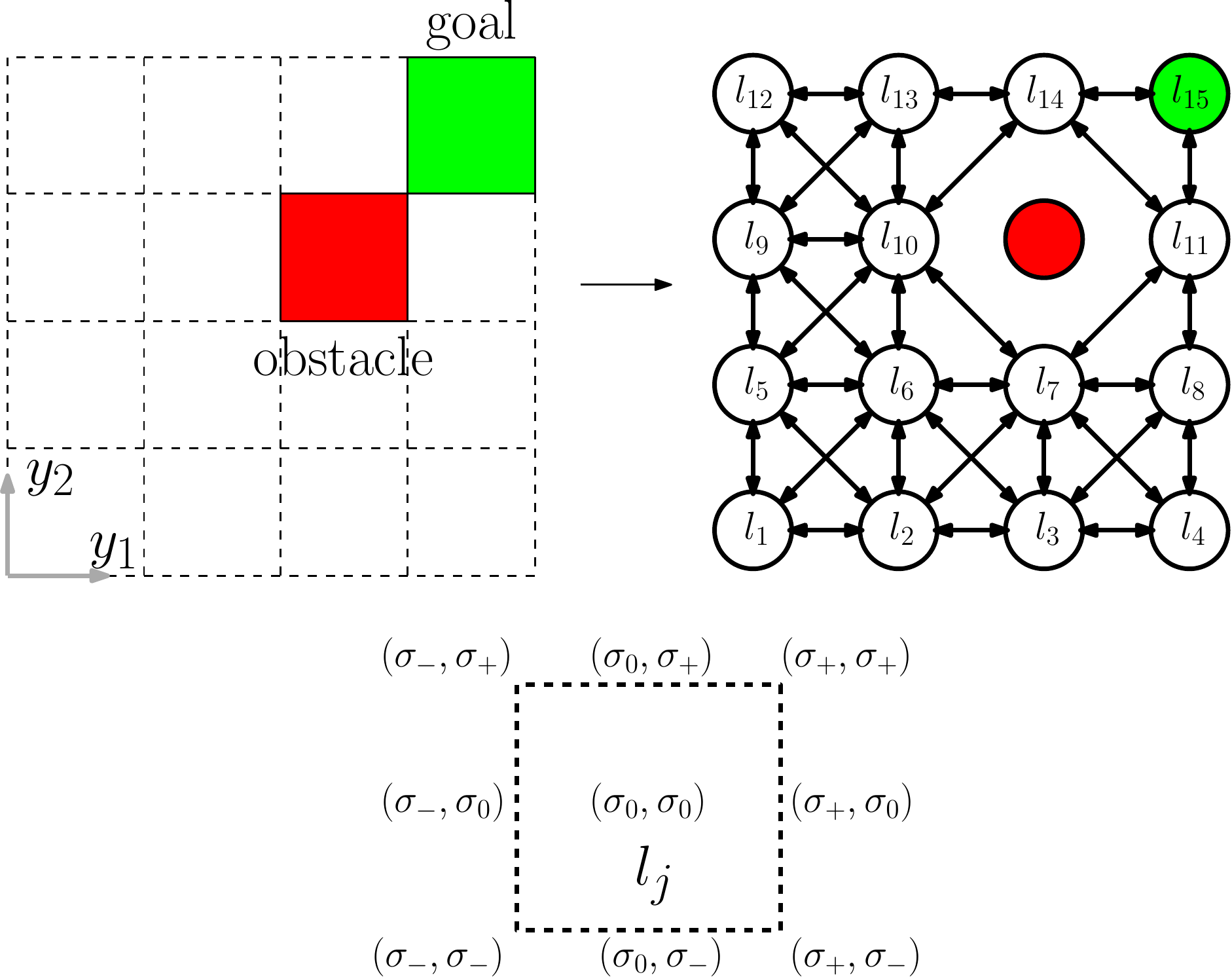}
\caption{A two-output, $p = 2$, example of a reach-avoid task. Shown on the left is the feasible space $\cP$ consisting of 16 boxes, the obstacle region $\cO$ (red), and the goal region $\cG$ (green). The output transition system (OTS), which abstracts the 15 non-obstacle box regions and their neighbor connectivity, is shown on the right. Shown below, the faces of a box are labelled using $\Sigma = \prod_{i=1}^2 \{\sigma_0, \sigma_+, \sigma_-\}$, which includes the box edges, vertices, and interior.}%
\vspace{-2mm}
\label{fig:OTSsquares}%
\end{figure}

The output transition system (OTS) provides an abstract description of the {\em workspace} or {\em output space} associated with the system \eqref{eq:nonlinsys}. Consider a canonical $p$-dimensional box $Y^* :=\prod_{i = 1}^p [0, d_i]$ with edge lengths $d_i > 0$. Let $\{Y_j \}_{j = 1}^{n_L}$ be a collection of boxes that covers the output space $\cP$. We assume that each box $Y_j$ is a translation of $Y^*$ by an integer multiple of $d_i$ in the $i$-th output coordinate. This yields a gridded output space as in Figure~\ref{fig:OTSsquares}. The OTS is a graph that abstracts this gridded space by associating a label $l_j$ to each box $Y_j$. Edges in the OTS encode contiguous boxes in the output space, where two boxes are said to be contiguous if they share any face of dimension between $1$ and $p-1$. Notice that the OTS does not describe the actual evolution of output trajectories of \eqref{eq:nonlinsys}.

The {\em output transition system} (OTS) is a tuple $\cAOTS = ( \LOTS, \Sigma, \EOTS )$ with the following components:

\begin{description}

\item[Locations] $\LOTS$ denotes a finite set of locations. Each location $l_j \in \LOTS$ is associated with a non-obstacle box $Y_j$ in the output space. 

\item[Labels] $\Sigma := \prod_{i=1}^p \{\sigma_0, \sigma_+, \sigma_-\}$ is a finite set of labels. Each label $\sigma \in \Sigma$ uniquely identifies the face of a box using the components $\sigma_0$, $\sigma_+$, $\sigma_-$. Refer to Figure \ref{fig:OTSsquares} for a $p = 2$ example.

\item[Edges] $\EOTS \subset \LOTS \times \Sigma \times \LOTS$ is a set of edges. The edge $e = (l, \sigma, l') \in \EOTS$ denotes that the face $\sigma$ of box $l$ is shared between $l$ and $l'$.

\end{description}

Figure \ref{fig:OTSsquares} shows an example OTS for a simple scenario. The OTS locations are the 15 non-obstacle boxes, which includes a goal box for the reach-avoid task. The OTS edges are shown as arrows; for example, $e = (l_6, (\sigma_-, \sigma_-), l_1) \in \EOTS$.

\subsection{Maneuver Automaton} \label{sec:MA}

The maneuver automaton (MA) is a hybrid system whose discrete states correspond to a finite set of motion primitives, and whose edges capture both constraints on feasible successive motion primitives and constraints on which faces of a box can be reached using a specific motion primitive. 
The concept of a MA was introduced in \cite{FRAZ05} as a way to simplify the control design by invoking canonical high-level behaviors. By concatenating motion primitives, complex behaviors such as reach-avoid can be achieved.

The {\em maneuver automaton} (MA) is a tuple $\cHMA = (\QMA,\Sigma,\EMA,\XMA,\IMA,\GMA,\RMA,\QMA^0) $, where

\begin{description}

\item[State Space]
$\QMA = M \times \RR^n$ is the hybrid state space, where $M$ is a finite index set
of {\em motion primitives}.

\item[Labels]
$\Sigma$ is the same set of labels used in the OTS, denoting the faces of a box.

\item[Edges]
$\EMA \subset M \times \Sigma \times M$ is a set of edges. The edge $e = (m, \sigma, m') \in \EMA$ denotes that the motion primitives $m$ and $m'$ may be executed in sequence by the system \eqref{eq:nonlinsys} when the 
output trajectory crosses the face $\sigma$ of the current box. 

\item[Vector Fields]
$\XMA$ assigns a closed-loop vector field of \eqref{eq:nonlinsys} for each motion primitive. Associated with this closed-loop vector field is a feedback controller that implements the desired motion primitive.

\item[Invariants]
$\IMA$ assigns an invariant region in the state space over which each motion primitive's feedback controller is defined. The projection of each invariant region onto the output space is contained in a box.

\item[Enabling and Reset Conditions]
Associated with each edge $e \in \EMA$ of the MA is an enabling condition $g_e \in \GMA$ that defines the set of states for which the edge can be taken. Likewise, the reset condition $r_e \in \RMA$ determines the new state value after the edge is taken.

\item[Initial Conditions]
$\QMA^0 \subset \QMA$ is a set of initial states.

\end{description}

\begin{remark} 
\label{rem:nondetMA}
Formally, an automaton is termed non-deterministic if there exists a state with more than one outgoing edge with the same label. In this way, the MA is non-deterministic as there are generally multiple selections for successive motion primitives. However, there is another relevant notion of non-determinism. Viewing only the discrete part of the MA, a given motion primitive may also have multiple edges with different labels, which has the interpretation that output trajectories can reach different faces non-deterministically. In contrast, in the full MA the face reached is a function of the initial condition.
\end{remark}

\subsection{Motion Primitives} 
\label{sec:MAexample}

\begin{figure}
\centering%
\includegraphics[width=0.6\linewidth,trim=0cm 0cm 0cm 0cm, clip=true]{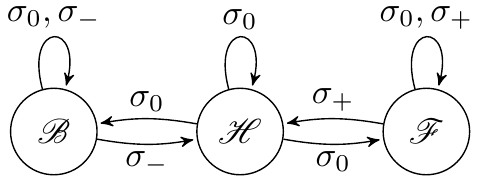}
\caption{A manuever automaton for double integrator dynamics, $p = 1$, with three motion primitives: Hold ($\sH$), Forward ($\sF$), and Backward ($\sB$). 
}%
\vspace{-2mm}
\label{fig:MATrans}%
\end{figure}

A motion primitive is a closed-loop vector field that achieves a desired canonical behavior in the output space. The closed-loop vector field of each motion primitive is formed by invoking a pre-computed feedback controller associated with that motion primitive and applying it to the system \eqref{eq:nonlinsys}.

Our design of motion primitives exploits three simplifications.
First, we invoke Assumption~\ref{assump:outputs} and the fact that all the boxes in the OTS are translations of each other to design motion primitives over the canonical box $Y^*$ only.
Second, we assume any feedback linearizations or transformations are done separately, in order to focus our design of motion primitives to double integrator dynamics in each output coordinate, noting that Assumption~\ref{assump:outputs} is satisfied.
Finally, to achieve an overall desired behavior in the $p$-dimensional output space, we first design {\em atomic motion primitives} for a single output coordinate and then compose them. Typical canonical motion primitives for robotic agents operating in a 2D workspace include {\em Right}, {\em Left}, {\em Up}, and {\em Down}. For example, if the desired motion primitive in the output space is Right, then we want an atomic motion primitive for $y_1$ to increase in value, and another atomic motion primitive for $y_2$ to hold its value.

\begin{figure}
\centering%
\includegraphics[width=1.0\linewidth,trim=3cm 2.3cm 2cm 1.0cm, clip=true]{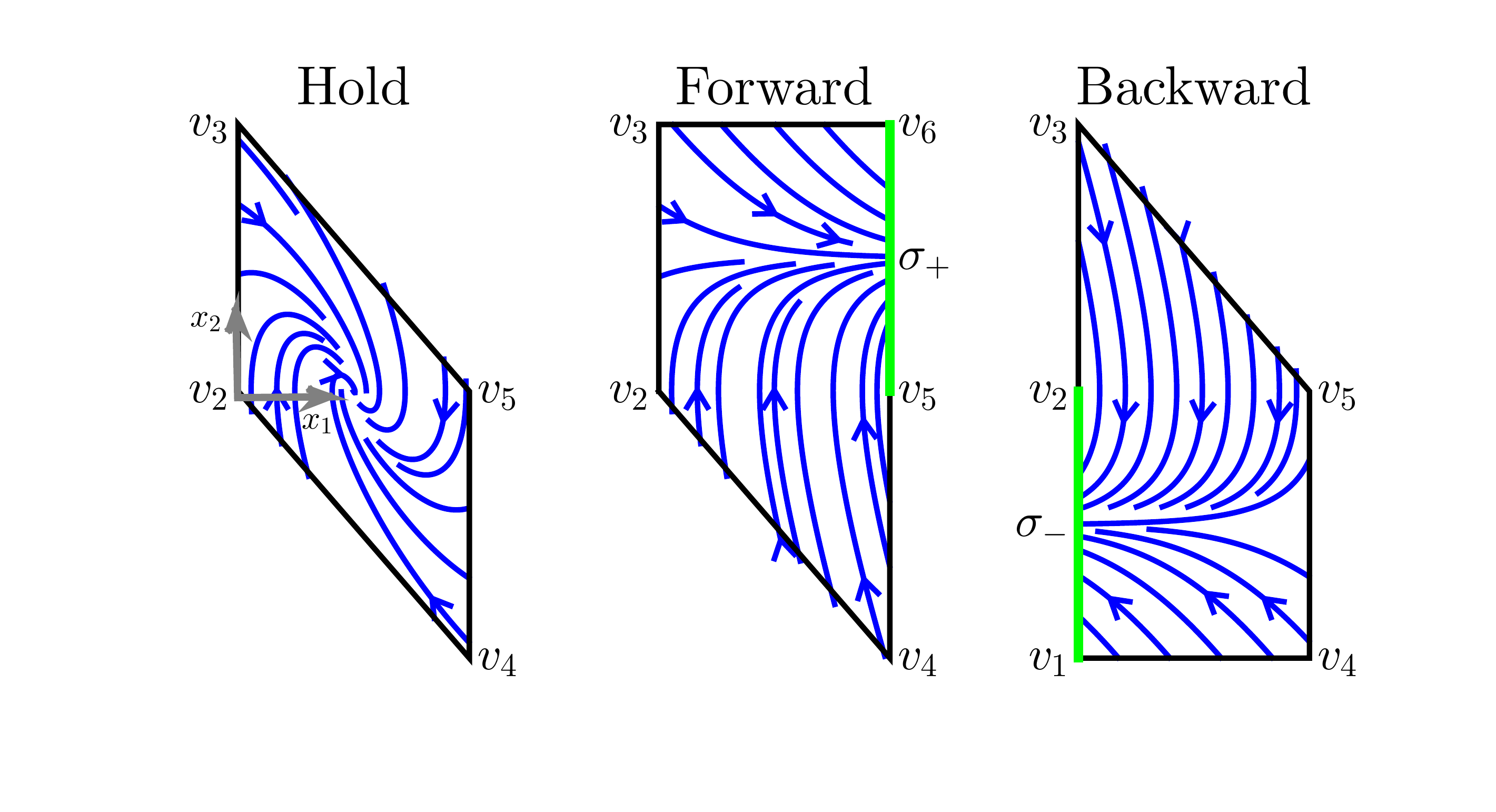} \\
\caption{The closed-loop vector fields for the Hold, Forward, and Backward motion primitives over their respective invariant regions in the $(x_1,x_2)$ axes.
}%
\vspace{-2mm}
\label{fig:AllModes}%
\end{figure}

Let \eqref{eq:nonlinsys} be the double integrator system
\begin{equation}
\dot{x}_1 = x_2, \;\; \dot{x}_2 = u, \;\;\;\;\; y = x_1, \; 
\end{equation}
where $x := (x_1,x_2) \in \RR^2$, $u \in \RR$, and the output $y$ is the position, so that $p = 1$. 
The MA, depicted in Figure~\ref{fig:MATrans}, consists of three atomic motion primitives, 
called {\em Hold}, {\em Forward}, and {\em Backward} and denoted as $\sH$, $\sF$, and $\sB$, respectively. Since $p = 1$, the canonical box $Y^*$ is simply a segment of fixed length. In the Forward motion primitive, output trajectories must exit the right face of $Y^*$ corresponding to the label $\sigma_+ \in \Sigma$, whereas in Backward it is the left face with $\sigma_- \in \Sigma$. For all three motion primitives, the label $\sigma_0 \in \Sigma$ corresponds to output trajectories having not yet crossed a face of $Y^*$. These behaviors are exhibited in the closed-loop vector fields shown in Figure \ref{fig:AllModes}. 

We synthesize controllers to achieve these closed-loop vector fields using reach control theory. In particular we employ standard design tools for reachability to a facet on a polytope for affine systems 
\cite{RB06,BRO13}. Each motion primitive's invariant region is a polytopic subset constructed as 
the convex hull of the vertices $v_k$, $k \in \{1, \ldots, 6\}$, see Figure \ref{fig:AllModes}.
The vertices are determined by the segment length $d > 0$, maximum control value $u^* > 0$, and the derived maximum speed $v^* = \sqrt{d u^*} > 0$.
Specifically, $v_1 = (0,-v^*)$, $v_2 = (0,0)$, $v_3 = (0,v^*)$, $v_4 = (d,-v^*)$, $v_5 = (d,0)$, and $v_6 = (d,v^*)$. 

For each $m \in \{ \sH, \sF, \sB \}$, we employ an affine feedback
\begin{equation}
\label{eq:controllaw}
u_m(x) = K_m x + g_m  \,.
\end{equation}
We set $K_{\sH} = \begin{bmatrix} k_1 & k_2 \end{bmatrix}$, 
$K_{\sF} = K_{\sB} = \begin{bmatrix} 0 & k_2 \end{bmatrix}$, $g_{\sH} = g_{\sF} = u^*$, $g_{\sB} = -u^*$, $k_1 := -2u^*/d$, and 
$k_2 := -2u^*/v^*$. 
These controllers can be derived by solving reach control problems on a triangulation of the polytopic subsets \cite{RB06,BRO13} and imposing continuity constraints.
We omit further details, but we refer to our previous work \cite{VUK16} for similar controllers.

In essence, these polytopic subsets are designed precisely so that the closed-loop vector fields over them can be pieced together and faithfully represent the discrete transitions defined in the MA shown in Figure \ref{fig:MATrans}. For example, the transition $(\sF, \sigma_+, \sH) \in \EMA$ is possible because when an output trajectory exits the right face of Forward as shown in green on Figure \ref{fig:AllModes}, it reappears on the left side of Hold (activating the enabling and reset conditions). The transitions labelled with $\sigma_0$ can be ignored when $p = 1$ and only play a role in correctly generating a MA for $p > 1$, which is described next.

The $p$-dimensional MA is obtained by parallel composing the single output MA shown in Figure \ref{fig:MATrans} $p$-times. Our three single-output motion primitives, $\{ \sH, \sF, \sB \}$, give rise to $3^p$ $p$-dimensional motion primitives, denoted as $\prod_{i=1}^p \{ \sH, \sF, \sB \}$. For example, $(\sF, \sH)$ would implement the motion primitive Right in 2D. The composed transitions are obtained by considering all possible combinations of one-dimensional transitions in each output. Consequently, this gives rise to non-deterministic motion primitives, such as $(\sF, \sF)$, but these may be optionally pruned out in the parallel composition.

\subsection{Product Automaton} \label{sec:product}

\begin{figure}
\centering%
\includegraphics[width=1\linewidth,trim=0cm 0cm 0cm 0cm, clip=true]{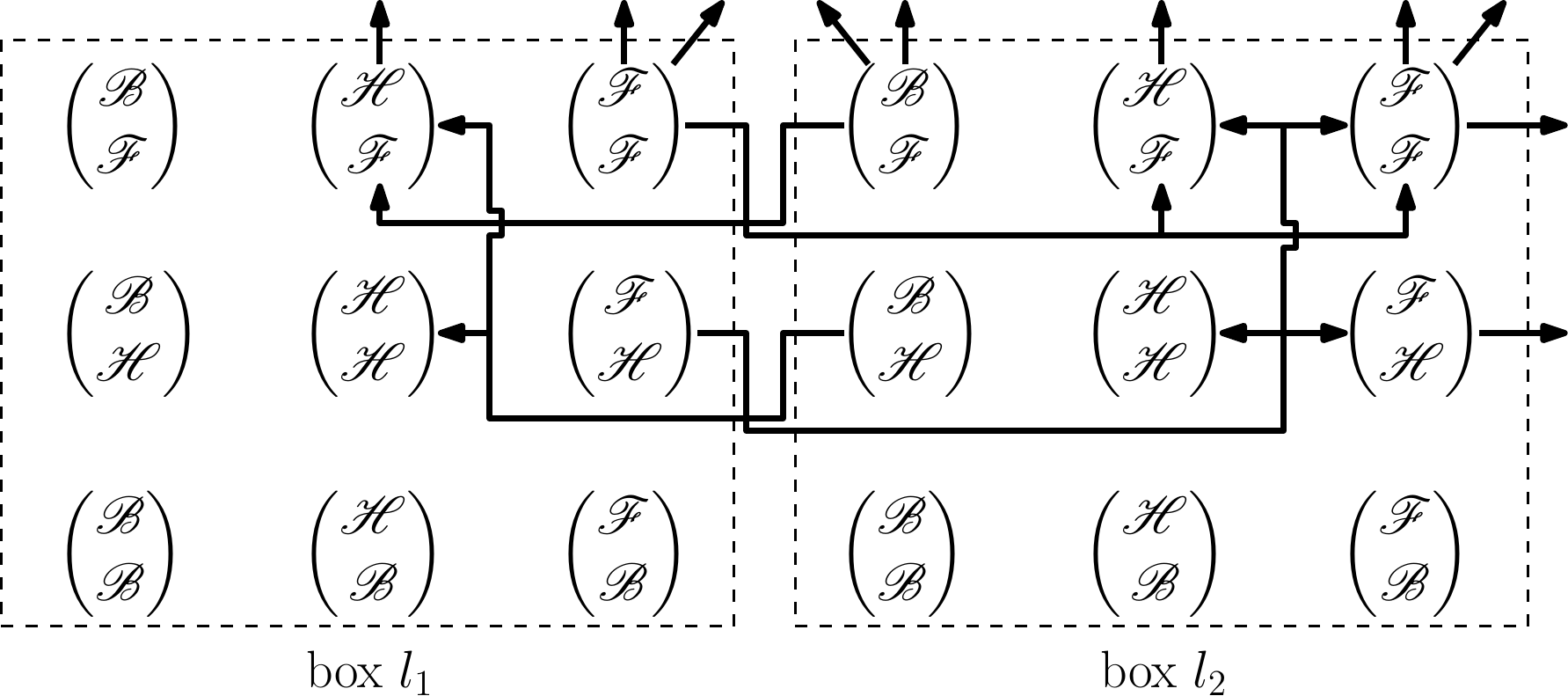}
\caption{A snapshot of the product automaton (PA) showing the system's motion capabilities over the two lower left boxes $l_1$ and $l_2$ from Figure~\ref{fig:OTSsquares}. The motion primitives shown in Figure \ref{fig:MATrans} are composed to obtain 9 motion primitives. A product state is a box equipped with a particular motion primitive. The arrow denotes an edge in the PA, where the outgoing location of the arrow corresponds to the face reached in a product state.}%
\vspace{-2mm}
\label{fig:product}%
\end{figure}

The product automaton (PA) is the synchronous product of the OTS and the discrete part of the MA. 
It models the combined constraints on successive locations in the OTS and successive motion primitives. As such, it captures the overall feasible motions of the system. Any high-level plan must adhere to these feasible motions.

We define the {\em product automaton} (PA) to be the tuple
$\cAPA = ( \QPA, \Sigma, \EPA, \QPA^f, \DPA, \HPA ) $, where

\begin{description}

\item[State Space]
$\QPA \subset \LOTS \times M $ is a finite set of states, denoting a box equipped with a motion primitive.

\item[Labels]
$\Sigma$ is the same set of labels used by the OTS and MA.

\item[Edges]
$\EPA \subset \QPA \times \Sigma \times \QPA$ is a set of edges. In particular, an edge 
$e = (q, \sigma, q') \in \EPA$, where $q = (l,m)$ and $q' = (l',m')$, if $(l,\sigma,l') \in \EOTS$ 
and $(m,\sigma,m') \in \EMA$.

\item[Final Condition]
$\QPA^f \subset \QPA$ is the set of final states. A product state $(l,m)$ is a final state if the box $l$ is labelled as a goal.

\item[Discrete and Terminal Cost]
$\DPA$ is the instantaneous cost associated with each edge in $\EPA$ and $\HPA$ is the terminal cost associated with each final state in $\QPA^f$.

\end{description}

Referring to Figure \ref{fig:product} for some examples, box $l_1$ equipped with $(\sF, \sH)$ is a product state, which will reach box $l_2$ in finite time through the face $(\sigma_+,\sigma_0)$. As can be verified, this product state has four edges each with the label $(\sigma_+,\sigma_0)$, corresponding to the feasible successive motion primitives $(\sH, \sH)$, $(\sF, \sH)$, $(\sH, \sF)$ and $(\sF, \sF)$. Each edge leads to a product state on the box $l_2$. On the other hand, box $l_1$ equipped with $(\sF, \sF)$ is another product state; however, only one of the three faces $(\sigma_+,\sigma_0)$, $(\sigma_0,\sigma_+)$, and $(\sigma_+,\sigma_+)$ will be reached in finite time. For each possibility there is a different set of edges. The PA inherits its non-deterministic properties from the MA, see Remark \ref{rem:nondetMA}.

\subsection{Hybrid Control Strategy}
Once the PA is constructed, we employ a variation of a standard non-deterministic Dijkstra algorithm \cite{BRO05, WOL13}. From this, we obtain a discrete feedback control policy, which assigns one successive motion primitive for each product state and for each possible face that can be reached from that product state. The hybrid control strategy consists of the discrete control policy and the low-level feedback controller associated with each motion primitive.

\begin{figure}
\centering%
\includegraphics[width=0.8\linewidth,trim=0cm 0cm 0cm 0cm, clip=true]{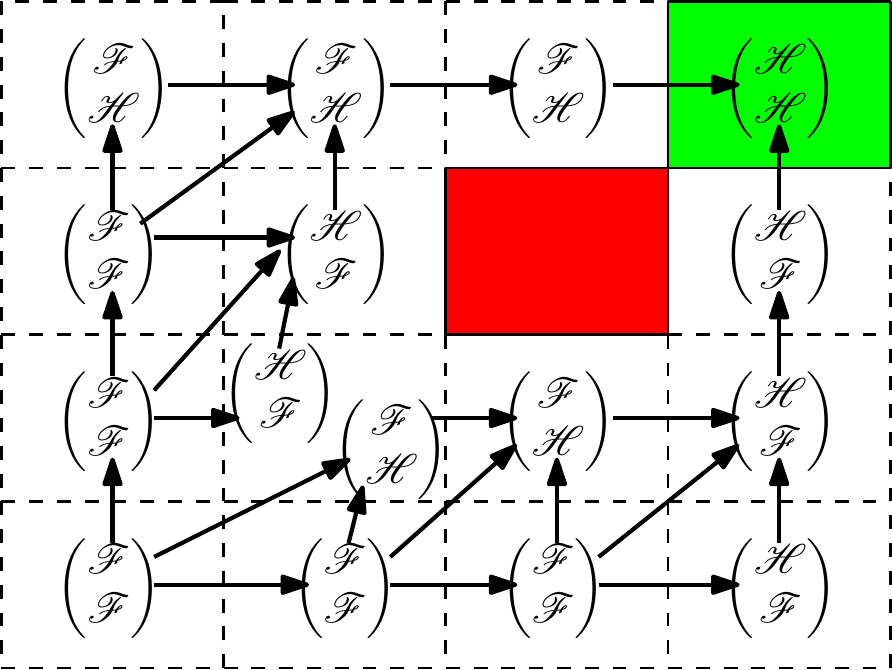}
\caption{This figure shows a discrete control policy for the scenario shown in Figure \ref{fig:OTSsquares}. }%
\vspace{-2mm}
\label{fig:ctrStrat}%
\end{figure}

Figure \ref{fig:ctrStrat} shows a possible control policy for the scenario from Figure \ref{fig:OTSsquares}. Although different routes may be taken from the same product state depending on the face reached, the control policy ensures that all paths lead to the goal.

\section{Quadrocopter Applications} \label{sec:application}

\subsection{Quadrocopter Modeling}
The standard quadrocopter dynamics model is ubiquitous in the literature; see, for example, \cite{LUP2014}. The vehicle dynamics are described by six degrees of freedom and are nonlinear. It is well known that this model is differentially flat \cite{SCHW14}. As a result, the dynamics for the position $(x_w, y_w, z_w)$ in the world frame each reduce to a double integrator and we are able to use the motion primitives from Section \ref{sec:MAexample}.

\subsection{Interfacing Multiple Quadrocopters}


To apply our framework to a joint reach-avoid task among $N$ quadrocopters, a copy of the gridded 3D workspace must be associated with each vehicle, resulting in a total of $p = 3N$ outputs. The $p$-dimensional MA representing the asynchronous motion capabilities of the multi-vehicle system is obtained by parallel composing the single-output MA from Section \ref{sec:MAexample}. We employ an exhaustive search over the multi-vehicle configurations to label the $p$-dimensional obstacle boxes, which accounts for real-world obstacles and pairwise collisions between any two vehicles.

\begin{figure}[t]
\centering%
\includegraphics[width=1\linewidth,trim=0cm 0cm 0cm 0cm, clip=false]{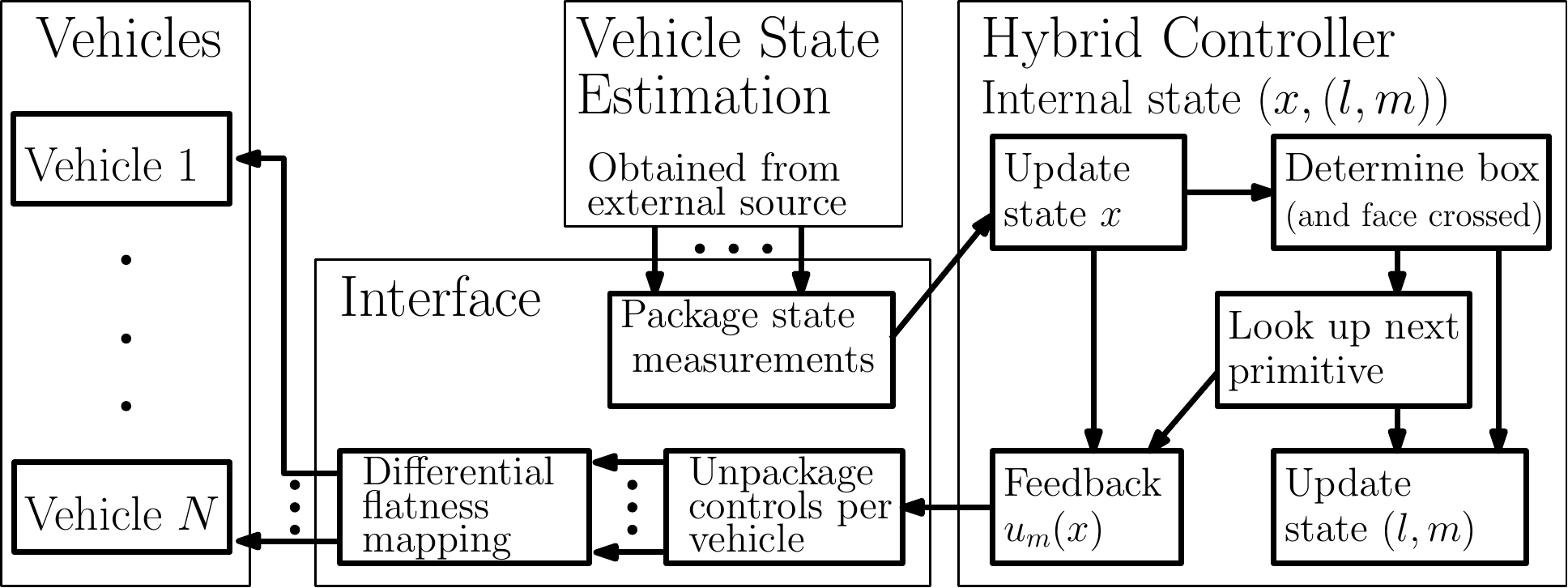}
\caption{Interface between multiple vehicles and the framework with $p = 3N$ outputs. The hybrid controller internal state consists of the joint state measurement of all the vehicles and includes the current (joint) box, $l$, and the current (joint) motion primitive, $m$. The internal state is updated via external state measurements (assumed to be given) and is used to compute the feedback controls.}%
\vspace{-2mm}
\label{fig:interface}%
\end{figure}

The multi-vehicle reach-avoid problem can be solved using our proposed methodology, following the steps shown in Figure \ref{fig:methodology}. The output of the methodology is a hybrid controller, which is fully computed offline. We discuss its computational complexity in Section \ref{sec:discussion}. Once the hybrid controller is computed, the system can successfully execute the reach-avoid task from any starting configuration corresponding to a valid initial condition of the hybrid controller. The runtime workflow is depicted in Figure \ref{fig:interface}, showing how the hybrid controller interfaces with the multi-vehicle system. 
Due to the simplicity of a box partition and assuming that the next motion primitive can be looked-up in constant time, each runtime component requires a negligible amount of computation, even for a large number of vehicles and outputs.

\subsection{Experimental Results}
\label{sec:experiment}

\begin{figure*}
\centering%
\includegraphics[width=0.32\linewidth,trim=0cm 0cm 0cm 0cm, clip=true]{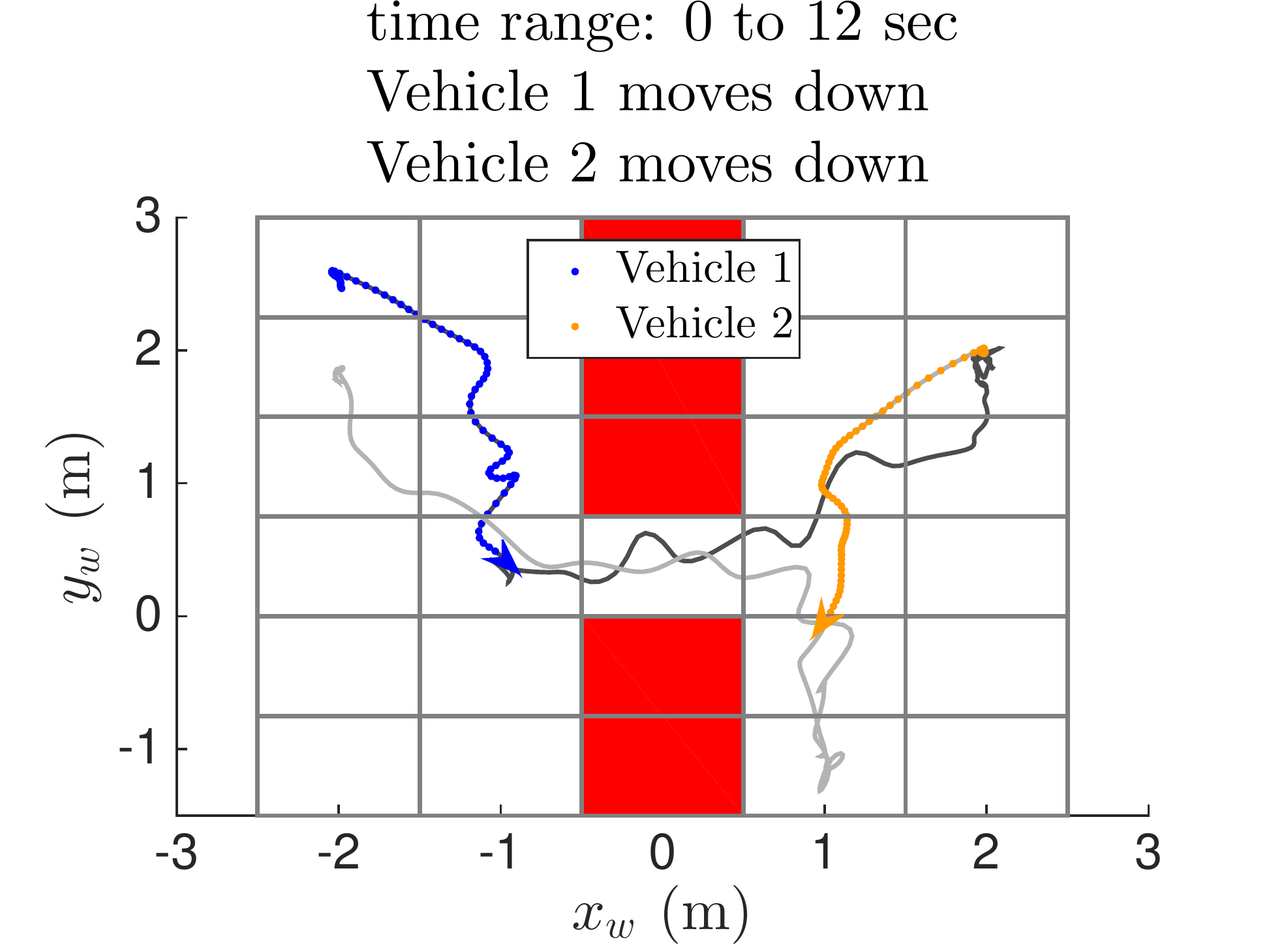}
\includegraphics[width=0.32\linewidth,trim=0cm 0cm 0cm 0cm, clip=true]{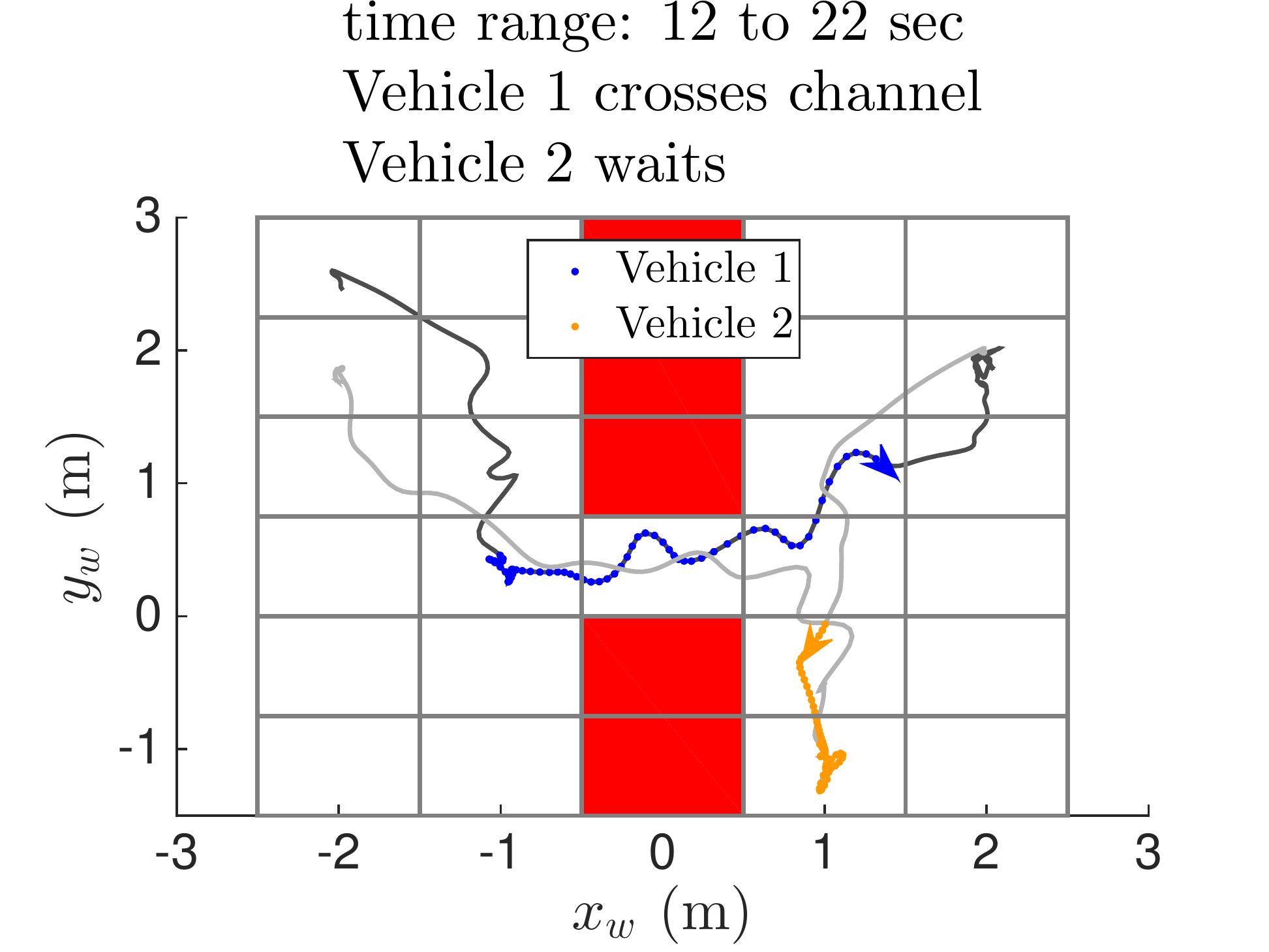}
\includegraphics[width=0.32\linewidth,trim=0cm 0cm 0cm 0cm, clip=true]{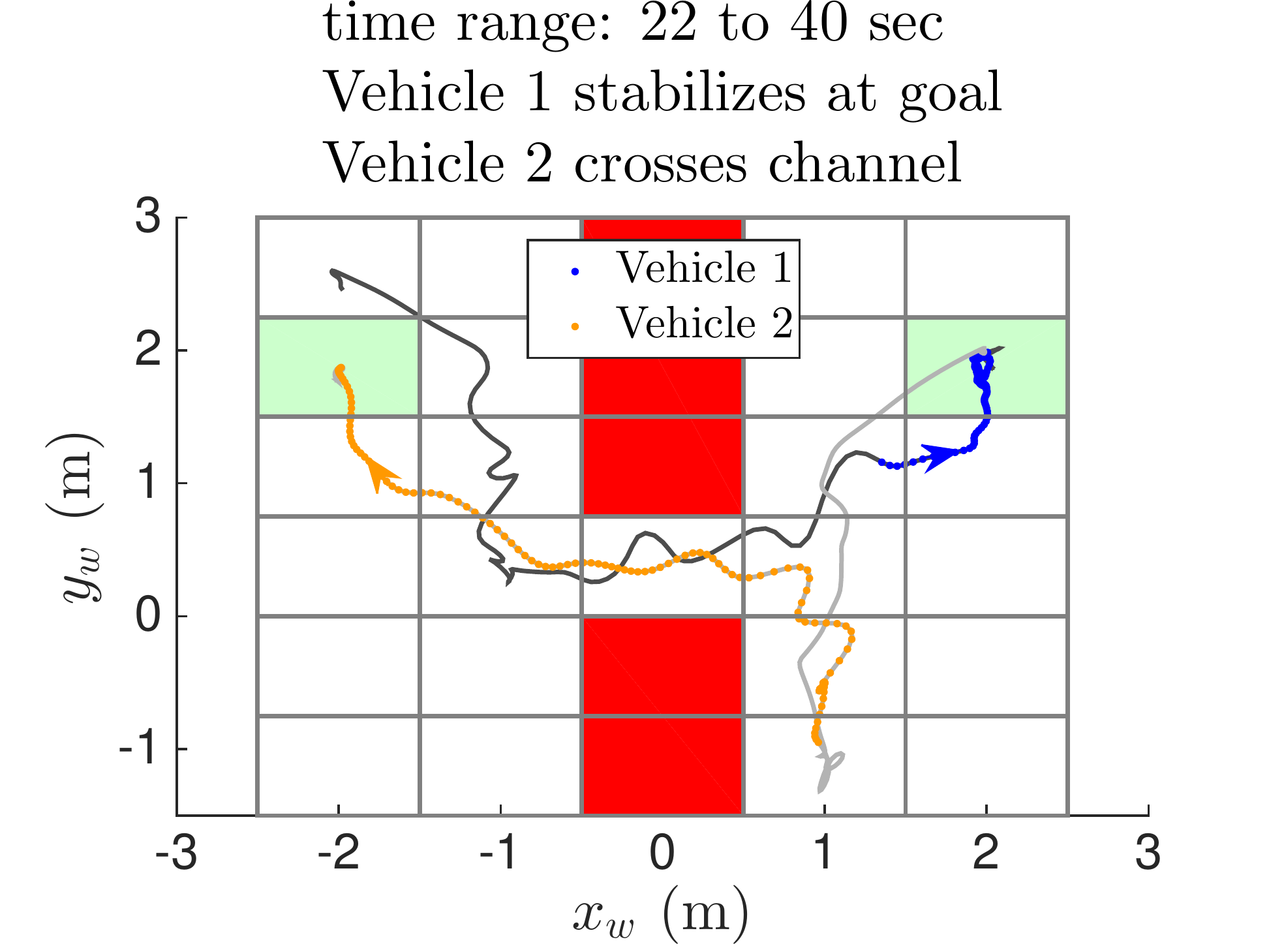}
\caption{Experimental results for the scenario where two quadrocopters must switch places through a narrow passage, where Vehicle 1 (blue) starts on the left and Vehicle 2 (orange) starts on the right. The grayed out trajectories show the full path followed by the vehicles; the color portion corresponds to the indicated time interval. The maneuver is shown over three time segments, with goals highlighted in green on the right. We observe that since Vehicle 2 gets to the passage first, it makes room and waits for Vehicle 1 to pass first. Noisiness in the trajectories is due to mutual aerodynamic effects, to be contrasted with Figure \ref{fig:nom1dr}. Although these effects are not accounted for, our hybrid feedback controller safely completes the task.}%
\vspace{-2mm}
\label{fig:xyDataNom}%
\end{figure*}

\begin{figure*}
\centering%
\includegraphics[width=0.32\linewidth,trim=0cm 0cm 0cm 0cm, clip=true]{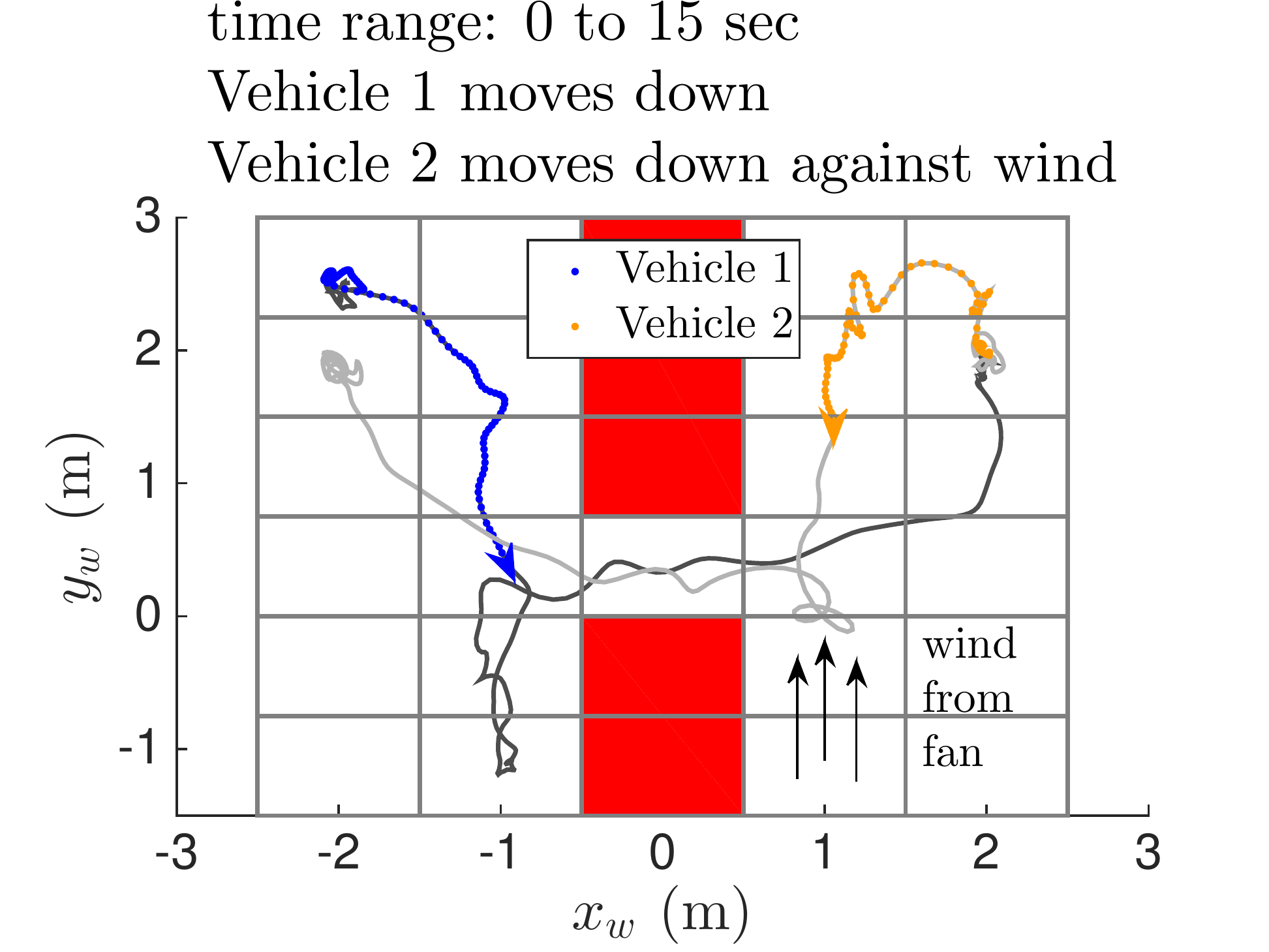}
\includegraphics[width=0.32\linewidth,trim=0cm 0cm 0cm 0cm, clip=true]{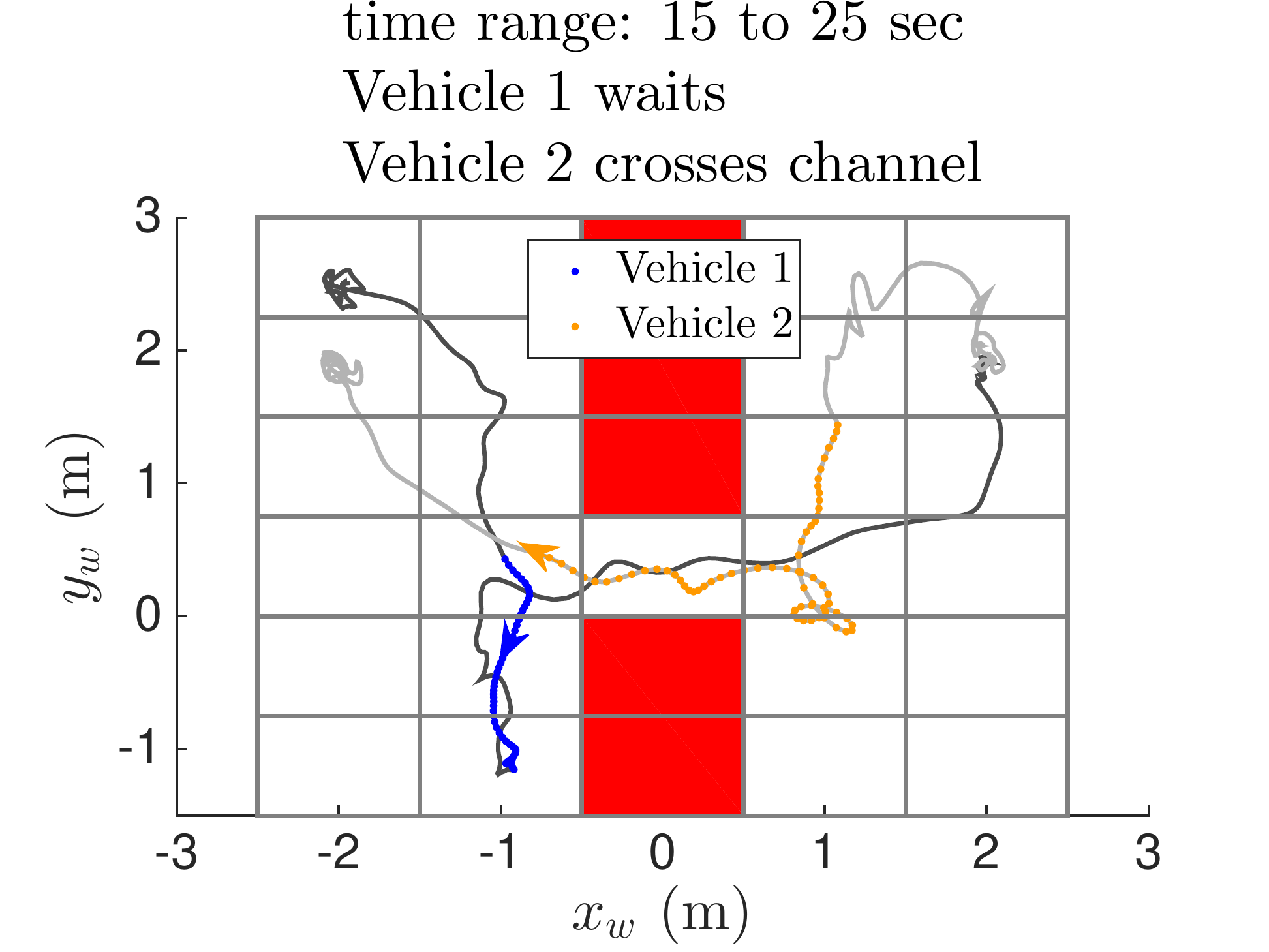}
\includegraphics[width=0.32\linewidth,trim=0cm 0cm 0cm 0cm, clip=true]{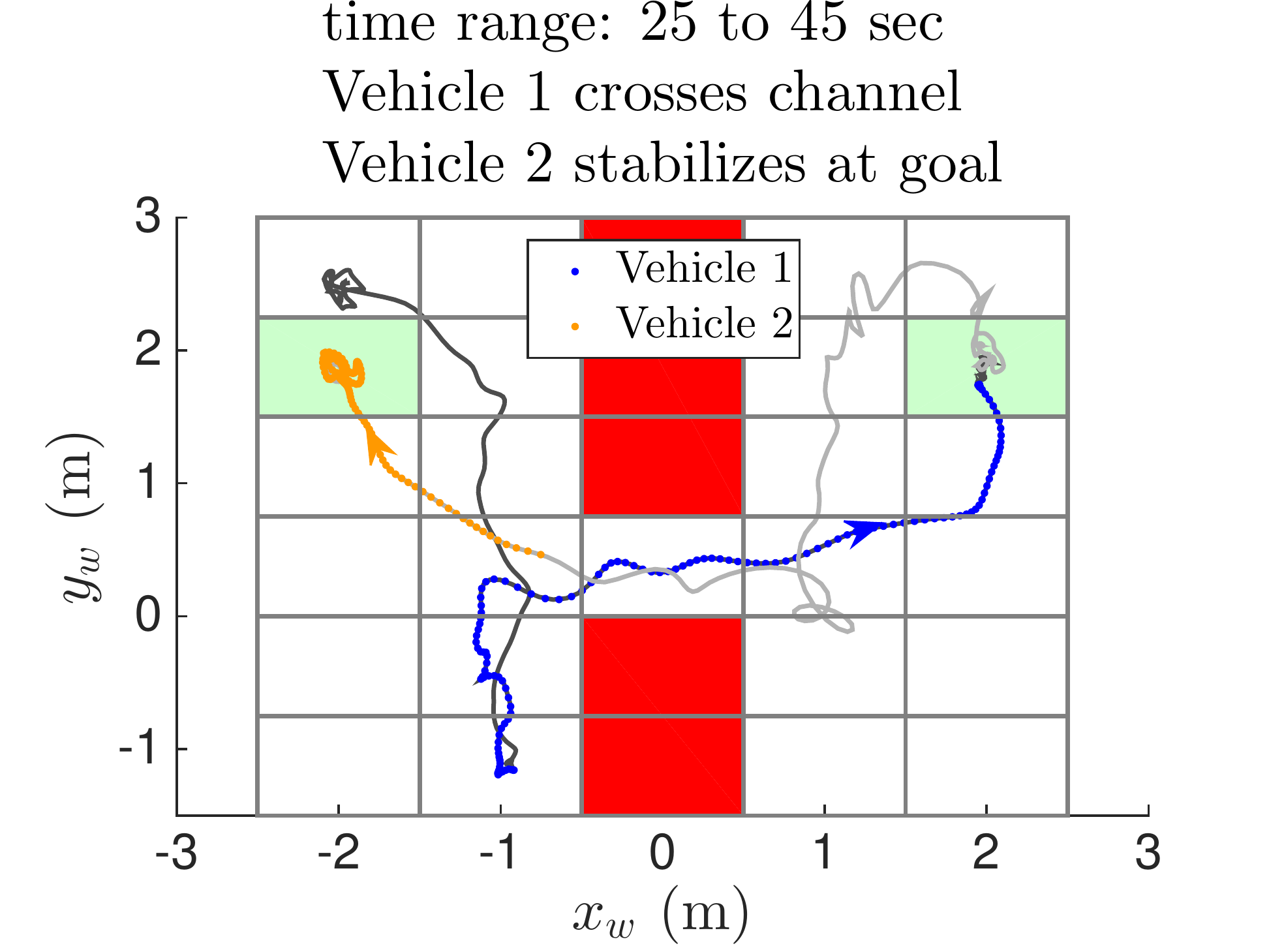}
\caption{The same scenario as in Figure \ref{fig:xyDataNom} above, but now an unmodelled wind disturbance causes a delay in Vehicle 2. To compensate, Vehicle 1 makes room and waits for Vehicle 2 to pass first, opposite to the nominal case. The same hybrid controller was used as in the nominal run of Figure \ref{fig:xyDataNom}.
}%
\vspace{-2mm}
\label{fig:xyDataDist}%
\end{figure*}

Our experimental platform is the Parrot AR.Drone 2.0 interfaced with the ROS \emph{ardrone autonomy} package. We used an external motion capture system to obtain the state estimates for our feedback controller \eqref{eq:controllaw}, which was run at 70 Hz. 

Due to limited space, we illustrate one interesting scenario where two quadrocopters must coordinate switching places through a narrow passage, see Figure \ref{fig:xyDataNom}.
Additional video examples include a simple, single quadrocopter maneuver, see \url{http://tiny.cc/quadrocopterPlanar}, and several 2D and 3D scenarios with two quadrocopters, see \url{http://tiny.cc/quad5scenes}.

For the example shown here, the 3D workspace is partitioned into a $(5 \times 6 \times 1)$ grid consisting of boxes having side lengths of 1, 0.75, and 3 meters, respectively.  The planar $(x_w,y_w)$-view is shown in Figure \ref{fig:xyDataNom}, where the red boxes represent the physical obstacles forming the passage. The $z_w$-direction is stabilized using the Hold motion primitive, so the overall number of outputs is effectively $p = 4$.

A nominal experimental run is shown in Figure \ref{fig:xyDataNom}, depicting the $(x_w,y_w)$-trajectories of each quadrocopter. The trajectories are divided into several time slots in order to show how the vehicles coordinate moving through the passage. 
As a consequence of our non-deterministic motion primitives, both vehicles move simultaneously for most of the run.
Notice that one vehicle (Vehicle 2 in this case) moves into a corner and waits to give space for the other vehicle to pass through the passage in order to maintain a sufficient safety margin.

\begin{figure}[t]
\centering%
\includegraphics[width=0.62\linewidth,trim=0cm 0cm 0cm 0cm, clip=false]{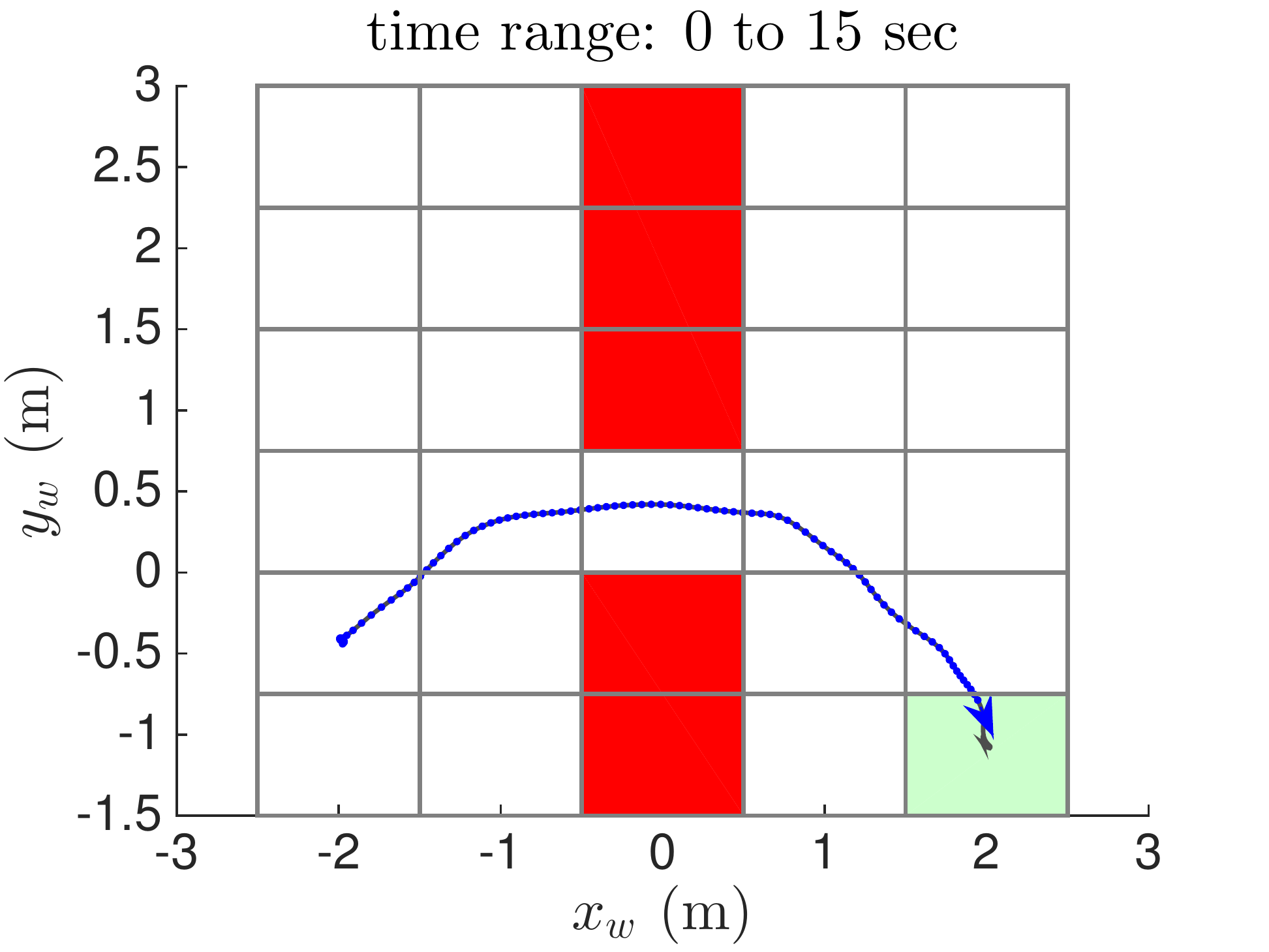}
\caption{A single-vehicle run for a similar scenario as considered in Figures \ref{fig:xyDataNom} and \ref{fig:xyDataDist}. The trajectory shown in blue is smooth and efficient, which illustrates that the noisy trajectories in the multi-vehicle case arise from mutually caused aerodynamic effects.}%
\vspace{-2mm}
\label{fig:nom1dr}%
\end{figure}

In contrast, an experimental run with disturbances induced by a fan is shown in Figure \ref{fig:xyDataDist}. 
Starting from the same configuration as the nominal run, Vehicle 2's motion was delayed due to a persistent wind disturbance, see the annotation. As such, the quadrocopters automatically executed a significantly different path compared to the nominal run (Vehicle 1 moved into a corner this time). In Figures \ref{fig:xyDataNom} and \ref{fig:xyDataDist} mutual aerodynamic effects resulted in ``wobbly" trajectories, which can be contrasted to the single vehicle run in Figure~\ref{fig:nom1dr}.

As long as disturbances are not too severe, the underlying feedback-based motion primitives will compensate for it and, if necessary, the high-level control policy will guide the vehicles along a new path to the goal configuration. Thus, our hybrid control strategy achieves a robust maneuver that does not require any online recomputation or timing estimates in the trajectories.

\section{Discussion} \label{sec:discussion}

\begin{figure}[t]
\centering%
\includegraphics[width=1\linewidth,trim=0cm 0cm 0cm 0cm, clip=true]{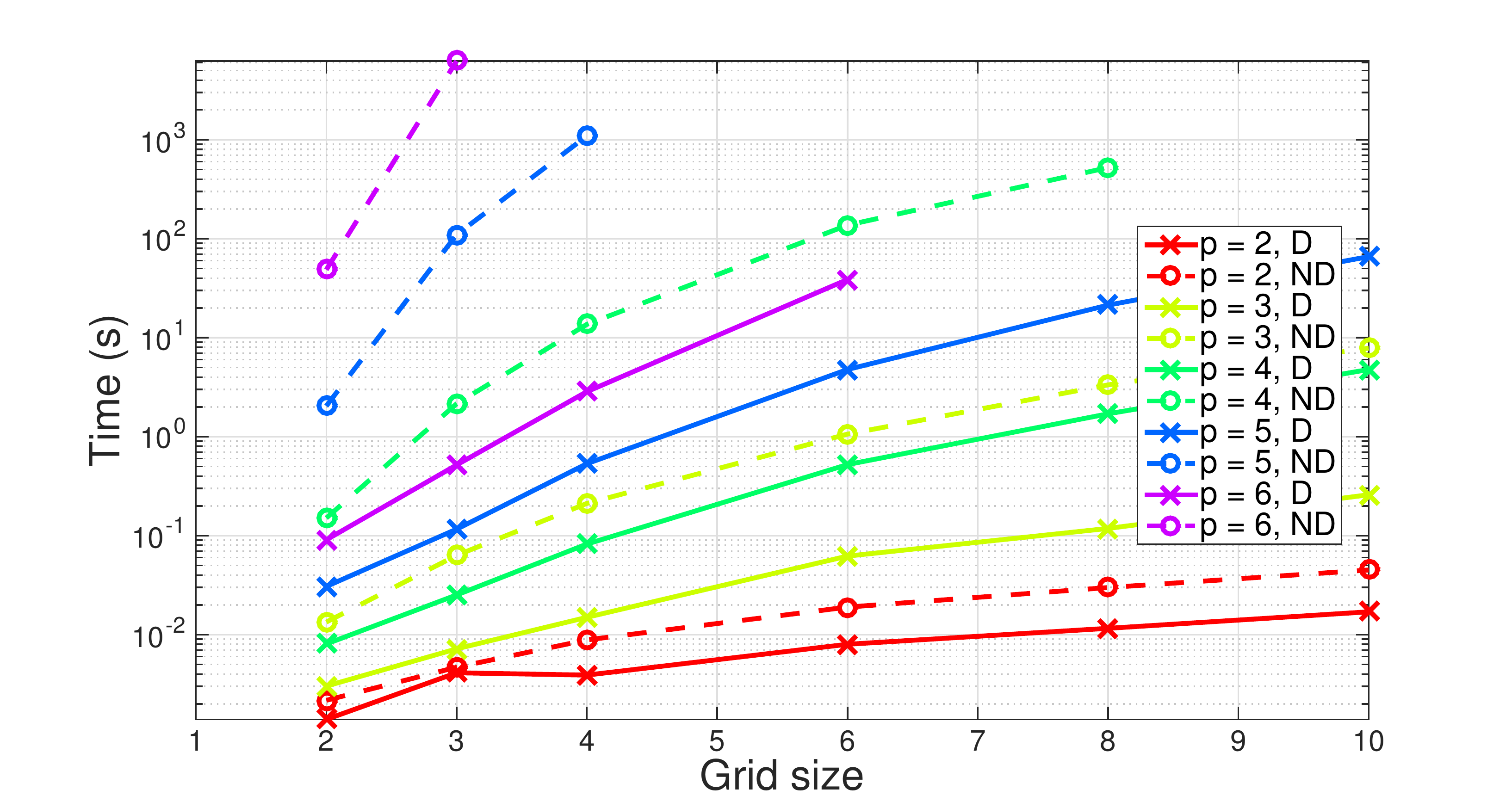} \\
\caption{Offline computation times for various scenario sizes. The grid size refers to the number of boxes in each output; for example, if $p = 3$ and the grid size is 4, there are a total of $4^3$ OTS boxes. We compare the computation using non-deterministic (ND) versus deterministic (D) motion primitives.}%
\vspace{-2mm}
\label{fig:complexity}%
\end{figure}

We analyze the complexity of the computations associated with our methodology. Using the MA shown in Figure \ref{fig:MATrans} for each output, Figure  \ref{fig:complexity} shows the total time to compute the OTS (with one goal and no obstacles), the parallel composed MA (with and without non-deterministic motion primitives), the PA, and a control policy (employing a non-deterministic Dijkstra algorithm) for various scenario sizes. Our implementation was done in Python 2.7.10 and computations were performed on a 64-bit Lenovo ThinkPad with a 2.6 GHz Intel Core i7 processor and 15.3 GiB RAM.



From Figure \ref{fig:complexity}, it is evident that the offline computation time is mild for moderate problem sizes. For small problems involving one or two vehicles, the computation often takes under a second, and all cases shown take less than two hours. For the scenario shown in Section \ref{sec:experiment}, the total computation time was about 8 seconds. A significant reduction in computation time is possible by restricting to use only deterministic motion primitives, at the expense of limiting simultaneous motion capabilities. Finally, since substantial motion can be robustly achieved through a single box, often the grid size can be quite coarse.

While it may be expensive to compute the full solution to a larger multi-vehicle system, it only needs to be performed offline once. Therefore, it may be an effective way to coordinate multiple vehicles through critical areas. Alternatively, the small computation times for $p \leq 3$ suggest a decentralized approach, in which each agent treats the other vehicles as obstacles and recomputes its hybrid control strategy over a small neighbourhood of boxes. Finally, faster computation times may be possible through specialized shortest path algorithms that do not explore the full product automaton structure.

\section{Conclusion} \label{sec:conclusion}
We have developed a modular, hierarchical framework for motion planning of heterogenous agents in known environments. It consists of various modules that describe the partitioned workspace, low-level control design, and high-level motion plan.
Overall we obtain a two-level control design which is highly robust, modular, and conceptually elegant. We presented a specific maneuver automaton for a double integrator system, and we illustrated its effectiveness to coordinate a maneuver between two quadrocopters. Extensions of this work include a decentralized implementation, a specialized shortest path algorithm to better exploit the modularity of our framework, and a generalization to control specifications expressed in linear temporal logic.

\bibliographystyle{IEEEtranS}

\end{document}